\documentstyle[preprint,aps,bezier]{revtex}

\begin{document}

\input FEYNMAN

\title{Becchi-Rouet-Stora-Tyutin quantization of a soliton model in 2+1
dimensions}

\author{
Juan P.\ Garrahan,$^{a,}$\thanks{E-mail:
garrahan@tandar.cnea.edu.ar.}$^{,}$\thanks{Fellow of the CONICET,
Buenos Aires, Argentina.} Luis M.\ Kruczenski,$^a$ Carlos L.\
Schat,$^a$ Daniel R.\ Bes,$^{a,\dagger}$ and Norberto
N.\ Scoccola$^{a,b,\dagger}$}

\address{$^a$ Departamento de F\'{\i}sica, CNEA, Av.\ Libertador 8250,
1429 Buenos Aires, Argentina.\\
$^b$ INFN, Sez.\ Milano, Via Celoria 16, 20133 Milano, Italy.}

\date{\today}
\preprint{TAN-FNT-94/10}

\maketitle

\begin{abstract}
The Becchi-Rouet-Stora-Tyutin (BRST) method is applied to the
quantization of the solitons of the non-linear $O(3)$ model in $2+1$
dimensions. We show that this method allows for a simple and systematic
treatment of zero-modes with a non-commuting algebra. We obtain the
expression of the BRST hamiltonian and show that the residual
interaction can be perturbatively treated in an IR-divergence-free
way. As an application of the formalism we explicitly evaluate the
two-loop correction to the soliton mass.
\end{abstract}


\section{Introduction}

During the last few years the issue of soliton quantization has
received renewed attention. One of the main reasons for this was the
revival of soliton models as the low energy limit of QCD \cite{Wit79}.
In this kind of models, baryons are described as solitonic excitations
of a chiral lagrangian in $3 + 1$ dimensions \cite{Sky61}.  Although
many aspects of the soliton quantization problem have been studied in
the mid-seventies \cite{GJ75,GS75,CL75,Tom75} the methods developed at
that time  become quite cumbersome when applied to systems in more than
one spatial dimension\footnote{Recently the problem of soliton
quantization has been addressed in Ref.\ \cite{CKW93} within the
framework of the Kerman-Klein method.}. As an alternative to these
canonical quantization methods one can apply the
Becchi-Rouet-Stora-Tyutin (BRST) quantization formalism \cite{HT92}.
This application \cite{BK90} is based on the observation that there
appears a fundamental gauge symmetry when collective coordinates are
introduced to perform the quantization of the zero modes around the
soliton.  Such gauge symmetry consists of the group of all
time-dependent transformations which simultaneously move the intrinsic
frame and the soliton so as to reproduce the same physical situation.
The overcompleteness associated with the introduction of collective
coordinates has to be compensated by constraints and gauge conditions.
In the BRST treatment both the collective coordinates and the Lagrange
multipliers associated with the constraints are considered as dynamical
variables on the same footing as the original coordinates.  Ghosts
(i.e., fermions carrying zero spin and having no direct physical
meaning) are introduced from the beginning. As an illustration of this
method, the authors of Ref.\ \cite{ABS92} have applied it to the
quantization of soliton models in $1+1$ dimensions, where comparison
with conventional methods can be easily performed.

However, it is not in such kind of problems where the advantage of
using the BRST scheme is more clearly seen (except perhaps for the fact
that solitons with arbitrary large momentum can be easily quantized
\cite{ABS93}). In the present paper we deal with the non-linear $O(3)$
model in $2+1$ dimensions \cite{Raj84}.  This model has already all the
complexity associated with the presence of several zero modes with a
non-commutative algebra, with the interplay between internal (isospin)
and external transformations and with the survival of some unbroken
symmetries.  At the same time, its rather simple form allows for
explicit expressions for the soliton configurations and zero-modes.

It should be stressed that this model has relevant applications on its
own. For example, in solid state physics it has been used as a model
for the continuum limit of a two-dimensional isotropic ferromagnet. In
that context the soliton solutions represent the metastable
`pseudo-particles' which are responsible for the destruction of
long-range order at any temperature \cite{BP75}. More recently, the
non-linear $O(3)$ model has also been studied in connection to
high-T$_c$ superconductivity \cite{DPW88}.

This paper is organized as follows. In Sec.\ \ref{SOL} we review the
non-linear $O(3)$ model in $2+1$ dimensions and present its soliton
solutions. In Sec.\ \ref{CCC} we introduce the collective coordinates
and impose the corresponding constraints. In Sec.\ \ref{Q2} the
quadratic hamiltonian is obtained and diagonalized. In Sec.\ \ref{BRST}
we describe the BRST quantization of the model. In Sec.\ \ref{DIAGS} we
show explicitly that the two loop correction to the soliton mass is
independent of the spurious parameters introduced by the BRST
quantization.  Finally, in Sec.\ \ref{CON} our conclusions are given.

\section{The model and its soliton solutions} \label{SOL}

The non-linear $O(3)$ model consists of three real scalar
fields $\phi_a$ ($a=1,2,3$) subject to the constraint
\begin{equation}
\phi_a \phi_a =1. \label{const}
\end{equation}
The dynamics is determined by the lagrangian density ($\mu = 0,1,2$)
\begin{equation}
{\cal L} = \frac{f^2}{2}\partial_{\mu}\phi_a\partial^{\mu}\phi_a,
	\label{lag}
\end{equation}
which is invariant under spatial translations and rotations and under
internal $O(3)$ transformations. As usual we take $\hbar=c=1$ and
work in terms of adimensional quantities. The constant $f$ is taken as
the large parameter of the model.  In the case of the
antiferromagnetism, $f$ corresponds to the spin of the ions which is
assumed to be large in the usual quasi-classical expansion
\cite{DPW88}. In the standard Skyrme model the role of $f$ is played by
the pion decay constant $f_\pi$ which is of ${\cal O}(N_c^{1/2})$ in
the large-$N_c$ expansion, and therefore a large number.

As it is well-known, the $O(3)$ symmetry is spontaneously broken by the
vacuum which we choose to be $\vec{\phi}=(0,0,1)$. Besides this
solution the equations of motion that result from the minimization of
the lagrangian (\ref{lag}) in the presence of the constraint
(\ref{const}) admit non--trivial, finite-energy, static solutions
satisfying $\lim_{|\vec{x}|\rightarrow\infty}
\vec{\phi}=(0,0,1)$~\cite{BP75}.  They represent  mappings of
$S_2^{(phy)}$ into $S_2^{(int)}$ and can be characterized by the
winding number W \footnote{Here and in what follows the surface element
is omitted in the space integrals.}
\begin{equation}
W = {1\over{8\pi}} \int \epsilon^{0ij} \epsilon^{abc}
	\phi_a \partial_j \phi_b \partial_i \phi_c.
\end{equation}

In order to obtain the explicit form of these soliton solutions it is
convenient to introduce independent fields. As shown in
Ref.\cite{Raj84} a useful choice is obtained by stereographically
projecting $S_2^{(int)}$ onto a  plane parallel to the
$\{\phi_1,\phi_2\}$ plane which contains the south pole.  The
(internal) cartesian coordinates thus obtained are used to construct
the complex fields
\begin{equation}
\theta = \frac{\phi_1 + i \phi_2}{1 - \phi_3} \; ; \;\;\;\;\;
	\bar{\theta} = \frac{\phi_1 - i \phi_2}{1 - \phi_3}.
\end{equation}
In the same fashion, the coordinates of the two dimensional physical
space $x^1,x^2$ are combined into independent complex coordinates
\begin{equation}
z= x^1+i\ x^2 \; ; \;\;\;\;\; {\bar z}=x^1-i \ x^2.
\end{equation}
In terms of the new fields the lagrangian density reads
\begin{equation}
{\cal L} = \frac{2 f^2}{(1+\theta \bar{\theta})^2}
	\left[
	\dot{\theta} \dot{\bar{\theta}} -
	2 (\partial_z \theta \partial_{\bar{z}} \bar{\theta} +
	   \partial_{\bar{z}} \theta \partial_z \bar{\theta})
	\right],
\end{equation}
where
\begin{equation}
\partial_z=\frac{1}{2}(\partial_1-i\partial_2) \; ; \;\;\;\;\;
	\partial_{\bar z}=\frac{1}{2}(\partial_1+i\partial_2).
\label{ky2}
\end{equation}
The equations of motion can be now easily obtained by applying the
Euler-Lagrange variational principle. For time-independent
configurations we obtain
\begin{equation}
\frac{2 \bar{\theta}}{1+\theta \bar{\theta}}
	\partial_z \theta \partial_{\bar{z}} \theta -
	\partial_{z {\bar{z}}} \theta = 0,
	\label{claseq}
\end{equation}
and similarly for ${\bar \theta}$.  It easy to see that any static
analytic $\theta(z)$ (or antianalytic $\theta(\bar{z})$) function
automatically solves Eq.\ (\ref{claseq}).  Since cuts are prohibited
because of the single-valuedness of $\phi_a(x)$, the only allowed
singularities of $\theta(z)$ are poles.  As shown in Ref.\ \cite{BP75},
the general form of the soliton solution is therefore a quotient of
polynomials. The corresponding winding number is given by the maximum
between the degrees of the numerator and denominator.  For winding
number $W=1$ the general solution satisfying $\theta\rightarrow\infty$
$(|x|\rightarrow\infty )$ is therefore the four-parameter family given
by
\begin{equation}
\theta(z)=az+b, \label{winding1}
\end{equation}
where $a,b$ are arbitrary complex constants. As a particular $W=1$
soliton solution, we choose $\theta=z$ ($a=1$, $b=0$).  Since the
soliton mass is independent of the values of the parameters $a,b$,
there will be four zero-modes around this solution. They correspond to
the transformations that connect our particular solution with any other
given by Eq.\ (\ref{winding1}).  As it will become clear below two of
these zero-modes are associated with spatial translations, one with
spatial and internal (isospin) rotations\footnote{There are three
independent isospin rotations in the $O(3)$ model. In fact, the
rotations around the 1 and 2 axes would yield zero frequency
(Goldstone) bosons already for the $W=0$ sector, due to the chosen
asymptotic vacuum $(0,0,1)$. In the $W=1$ sector the existence of such
modes becomes apparent if we use instead of (\ref{winding1}) the more
general six-parameter family $\theta(z)=(az+b)/(cz+d)$ subject to the
condition $ad-bc=1$. For the two remaining isospin modes we assume that
they can be IR-regularized  including a (small) mass term for the
Goldstone bosons, in both the $W=0,1$ sectors.} and the remaining one
with spatial dilatation.

\section{Collective coordinates and constraints} \label{CCC}

In general, all the zero-modes are associated with symmetries of the
action which are broken by the classical solution.  An exception to
this are the dilatations. In fact, the action is not invariant under
dilatations (although the classical energy is). In addition, in more
realistic models, where the effective lagrangian includes terms with
higher powers in the field  derivatives, the dilatation does not
correspond to a zero-mode at all \cite{Veg78}.  Considering this
situation and since we are mainly interested in the treatment of
collective variables associated with the breakdown of symmetries, we
will ignore the motion associated with dilatations altogether.

In order to treat the remaining zero-modes of the system  we include
collective coordinates and describe the system from a rotated and
translated spatial frame. The position of the moving frame with respect
to the fixed (laboratory) frame is determined by the collective
variables:  the displacements $Z$ and $\bar{Z}$ for the translation and
the angle $\Phi$ for the rotation.

If $z',\bar{z}'$ stand for the coordinates in the laboratory frame, and
$z,\bar{z}$ for the coordinates in the moving frame, we have
\begin{equation}
z' \rightarrow z = e^{-i \Phi} (z'+ Z) \; ; \;\;\;\;\;
        \bar{z}' \rightarrow \bar{z} =
	e^{i \Phi} (\bar{z}'+ \bar{Z}).
	\label{ONA}
\end{equation}
The lagrangian density can be written as ($\mu = t,z,\bar{z}$)
\begin{equation}
{\cal L} = \frac{2 f^2}{(1+\theta\bar{\theta})^2}
	g^{\mu\nu}\partial_{\mu}\theta\partial_{\nu}\bar{\theta},
\end{equation}
where the metric tensor $g^{\mu \nu}$ is no longer constant in the
moving frame, but instead reads
\begin{equation}
g^{\mu \nu} =
	\left(
	\begin{array}{ccc}
	1 & \dot{z} & \dot{\bar{z}} \\
        \dot{z} & \dot{z}^2 & \dot{z} \dot{\bar{z}} - 2 \\
        \dot{\bar{z}} & \dot{z} \dot{\bar{z}} - 2 &
	\dot{\bar{z}}^2 \\
        \end{array}
        \right),
\end{equation}
where
\begin{eqnarray}
\dot{z} &=& v - i {\dot \Phi} z \; ; \;\;\;\;\;
        v = e^{-i \Phi} \dot{Z}, \nonumber \\
\dot{\bar{z}} &=& \bar{v} + i {\dot \Phi} \bar{z}
	\; ; \;\;\;\;\;
        \bar{v} = e^{i \Phi}  \dot{\bar{Z}}.
\label{rotcoor}
\end{eqnarray}
In order to include internal rotations we measure the fields from a
moving internal frame related to the fixed internal frame by an $U(1)$
transformation, i.e.,
\begin{equation}
\theta' \rightarrow \theta = e^{i\alpha} \theta'.
\end{equation}
The lagrangian density reads $(k=z,\bar{z})$
\begin{equation}
{\cal L} = \frac{2 f^2}{(1+\theta\bar{\theta})^2}
	\left[ (\dot{\theta}+i\dot{\alpha}\theta)
	(\dot{\bar{\theta}}-i\dot{\alpha}\bar{\theta})+g^{0k}
	\partial_k\bar{\theta}(\dot{\theta}+i\dot{\alpha}\theta) +
	g^{0k}\partial_k\theta(\dot{\bar{\theta}} -
	i\dot{\alpha}\bar{\theta}) \right].
\end{equation}

Since  we are considering the collective coordinates as true variables
of the problem, the independent degrees of freedom are the independent
fields $\theta$ and $\bar \theta$ {\em plus} the collective
coordinates. To compute the canonical hamiltonian we must first find
the conjugate momenta to these independent degrees of freedom. For the
case of the field $\theta$ we have
\begin{equation}
\pi = \frac{\partial {\cal L}}{\partial \dot{\theta}} =
      	\frac{2 f^2}{(1+\theta\bar{\theta})^2}
	\left( \dot{\bar{\theta}} - i\dot{\alpha}\bar{\theta}+
 	g^{0k}\partial_k\bar{\theta} \right).
	\label{pi}
\end{equation}
In a similar way $\bar \pi$ can be calculated by replacing $\theta$ by
$\bar \theta$ in Eq.\ (\ref{pi}). The Eqs.\ defining the conjugate
momenta to the collective variables yield the primary constraints
\begin{eqnarray}
T_3 &=& \frac{\partial L}{\partial \dot{\alpha}} = i\int
	\left( \pi \theta - \bar \pi \bar \theta \right)
	\equiv t_3, \nonumber \\
J &=& \frac{\partial L}{\partial {\dot \Phi}} =
        i \int
	\left[
	\pi ( \bar{z} \partial_{\bar{z}} - z \partial_z) \theta +
	\bar \pi ( \bar{z} \partial_{\bar{z}} - z \partial_z)
	\bar \theta \right]
	\equiv j, \nonumber       \\
P &=&\frac{\partial L}{\partial v} =
        \int
	( \pi \partial_z \theta + \bar \pi \partial_z \bar \theta )
	\equiv p, \nonumber \\
\bar{P} &=& \frac{\partial L}{\partial \bar{v}} =
    	\int ( \pi \partial_{\bar{z}} \theta +
        \bar \pi \partial_{\bar{z}} \bar \theta )
	\equiv{\bar p},
   	\label{ky3}
\end{eqnarray}
$T_3$ is the generator of collective internal rotations around the
$3$-axis and $J$, $P$ and $\bar{P}$ are the generators of collective
spatial rotations and translations. The operators $t_3$, $j$, $p$ and
$\bar{p}$ transform correspondingly the fields. The  external rotation
and the translations are associated with the Euclidean E(2) group
\begin{equation}
	[j,p]=-p \; ; \;\;\;\;\;
	[j,{\bar p}]= {\bar p} \; ; \;\;\;\;\;
	[p,{\bar p}]=0.
\end{equation}

For convenience we have defined $P$ and $\bar{P}$, which are the
generators of collective translations in a frame which is rotated in an
angle $\Phi$ with respect to the lab.\ frame (cf.\
Eq.\ (\ref{rotcoor})).  They do not commute with $J$. The corresponding
generators of translations parallel to the laboratory axes,
\begin{equation}
P_L = \frac{\partial L}{\partial \dot{Z}} = e^{-i \Phi} P
	\; ; \;\;\;\;\;
	\bar{P}_L = \frac{\partial L}{\partial \dot{\bar{Z}}}
	= e^{i \Phi} \bar{P},
	\label{pl}
\end{equation}
are, of course, the ones that commute with $J$.

It must be noted that the classical solution $\theta = z$ only
partially breaks spatial and isospin rotational symmetry. It is
invariant under the transformation generated by $t_3+j$. Therefore,
quantum excitations can be classified by the eigenvalues of this
operator. The collective variable conjugate to $T_3 + J$ is redundant.
The constraint $T_3 + J = t_3 + j$ insures that the operations
associated to $T_3 + J$ are determined by the intrinsic structure.

\section{The hamiltonian} \label{Q2}

As it is well known, the collective coordinates and momenta do not
appear in the expression of the canonical hamiltonian, namely
\begin{eqnarray}
H=\int {\cal H} &=&
        T_3 \dot{\alpha} + J {\dot \Phi} + P v + \bar{P} \bar{v} +
        \int
	\left( \pi \dot{\theta} + \bar{\pi} \dot{\bar{\theta}} \right) -
	\int {\cal L} \nonumber \\
        &=&
	\frac{1}{2f^2} \int (1+\theta {\bar \theta})^2\pi {\bar \pi} +
	\int \frac{4 f^2}{(1+\theta {\bar \theta})^2}
	\left( \partial_z \theta \partial_{\bar{z}} \bar{\theta} +
	\partial_{\bar{z}} \theta \partial_z \bar{\theta} \right).
	\label{kp1}
\end{eqnarray}
It follows immediately that the constraints are first-class.

In order to get the explicit form of the quadratic hamiltonian we
expand the fields and their conjugate momenta around the classical
solution as follows
\begin{equation}
\theta = z + \frac{(1+r^2)}{f} \hat{\theta} \; ; \;\;\;\;\;
	\pi = \frac{f}{1+r^2} \hat{\pi}.
\end{equation}

The quadratic hamiltonian density reads
\begin{eqnarray}
{\cal H}^{(2)} &=& \frac{1}{2} \hat{\pi} \hat{\bar{\pi}} -
	\frac{8(1-r^2)}{(1+r^2)^2}{\hat \theta}{\hat {\bar \theta}} +
	2 \partial_r{\hat \theta} \partial_r{\hat {\bar \theta}} +
	\frac{2}{r^2}\partial_{\varphi}{\hat \theta}
	\partial_{\varphi}{\hat {\bar \theta}} +
	\frac{8i}{1+r^2}
	{\hat {\bar \theta}}\partial_{\varphi} {\hat \theta}.
	\label{k49}
\end{eqnarray}

In order to diagonalize this hamiltonian it is convenient to use the
partial wave decomposition
\begin{equation}
{\hat \theta}= \sum_{nm} {a_{nm}\over{\sqrt{2\pi}}}
	R_{nm}(r) \exp{(im\varphi)},
\end{equation}
where $a_{nm}$ are complex numbers that satisfy $|a_{nm}|^2 = 1$, and
the real functions $R_{nm}$ are conveniently normalized. The resulting
eigenvalue equation reads
\begin{equation}
R_{nm}''(r) + \frac{1}{r}R_{nm}'(r) -
	\left(\frac{m^2}{r^2}-\frac{4m}{1+r^2}-
	\frac{4(1-r^2)}{(1+r^2)^2} \right) R_{nm}(r) =
	-\varepsilon^2_{nm}R_{nm}(r). \label{eve}
\end{equation}
As  expected, Eq.\  (\ref{eve}) has zero-energy solutions. In general,
they have the form \begin{equation} R_{0m} \propto {r^m\over{1+r^2}}.
\end{equation} It should be noticed, however, that those with $m > 1$
do not correspond to zero-energy modes around the $W=1$ soliton and
therefore have to be dismissed.  Combinations of the remaining four
zero-energy wavefunctions (two independent choices for each value of
$a_{0m}$ with $m=0,1$) describe the zero-modes mentioned in
Sec.\ \ref{SOL}. The explicit form of these linear combinations will be
given below.

In addition to the zero-energy solutions, Eq.\  (\ref{eve}) has a
continuum of finite energy solutions. The asymptotic forms of these
solutions read
\begin{equation}
\begin{array}{lcl}
R_{nm} \approx r^{|m|} & \mbox{for} & r\rightarrow 0, \\
R_{nm} \approx r^{-\frac{1}{2}} \sin{(\varepsilon_{nm}r+\delta_{nm})} &
	\mbox{for} & r \rightarrow \infty. \\
\end{array} \label{ccR}
\end{equation}
Using these boundary conditions, the eigenvalue Eqs.\  can be numerically
solved.  In this way the phase shifts $\delta_{nm}$ are obtained as a
function of the energies $\varepsilon_{nm}$.

The quadratic hamiltonian (\ref{k49}) commutes with the linear
expressions for the generators defined in (\ref{ky3})
\begin{eqnarray}
t_3^{(1)} &=& - j^{(1)} =
	i f\int \frac{z \hat{\pi} - \bar{z} \hat{\bar{\pi}}}{1+r^2},
	\label{kj23} \\
p^{(1)} &=&
	f\int \frac{\hat{\pi}}{1+r^2}, \nonumber \\
\bar{p}^{(1)} &=&
	f\int \frac{\hat{\bar{\pi}}}{1+r^2}. \nonumber
\end{eqnarray}

The operator $t_3^{(1)}+j^{(1)}$ vanishes due to the fact that the
symmetry is only partially broken. For this reason it is convenient
heron to use the following linear combinations of generators\footnote{A
primed $s$ will exclude $s=0'$.  Similar combinations and conventions
hold for the collective generators, which are denoted by $V_s$.}
\begin{eqnarray}
v_{s=0'}&=&\frac{1}{2}(t_3+j), \nonumber \\
v_{s=0}&=&\frac{1}{2}(t_3-j), \nonumber \\
v_{s=+1}&=& i \sqrt{2} \bar p, \nonumber \\
v_{s=-1}&=& - i \sqrt{2} p.
\end{eqnarray}

We may determine both the value of the inertia parameters $\Im_{s'}$
associated with the collective motion and the expression for  $G_{-s'}$,
the variables conjugate to the $v^{(1)}_{s'}$, through the well known
RPA Eqs.\  \cite{MW70}
\begin{equation}
{[H^{(2)},G_{s'}]}= -\frac{i}{\Im_{s'}}v^{(1)}_{s'} \; ; \;\;\;\;\;
	{[G_{s'},v^{(1)}_{-t'}]}=i\delta_{s't'}, \label{kg1}
\end{equation}
which yield
\begin{eqnarray}
G_0 &=& - \frac{2 i f}{\Im}
	\int \frac{\bar{z} \hat{\theta} - z \hat{\bar{\theta}}}{1+r^2},
	\nonumber\\
G_{-1} &=& - \frac{2 \sqrt{2} i f}{M}
	\int \frac{\hat{\bar{\theta}}}{1+r^2}, \nonumber\\
G_{+1} &=& \frac{2 \sqrt{2} i f}{M}
	\int \frac{\hat{\theta}}{1+r^2}, \nonumber \\
\Im_0 &=& \Im = 4 f^2
	\int \frac{r^2}{(1+r^2)^2}, \nonumber \\
\Im_{\pm 1} &=& M = 4 f^2
	\int \frac{1}{(1+r^2)^2}, \nonumber
\end{eqnarray}
where $G_{s'}={\cal O}(f^{-1})$ and $\Im_{s'}={\cal O}(f^2)$.

The integral defining $\Im$  diverges logarithmically. Therefore, we
will consider all space integrals up to an upper radius ${\cal R}$. The
limit ${\cal R} \rightarrow \infty$ can be taken safely at the end
since the corrections to the classical results are expressed as ratios
$(\Im_s\varepsilon_{nm})^{-1}$ between collective and intrinsic
energies.  In (more realistic) models displaying stability against
dilatations, the  rotational parameter would not vanish.

In terms of the creation and annihilation operators for the finite
frequency modes (see Eq.\ (\ref{modosn})) and of the linear generators,
the quadratic hamiltonian reads\footnote{As explained at the beginning
of Sec.\ \ref{CCC}, the dilatations have been ignored.}
\begin{eqnarray}
H^{(2)} & = & \varepsilon_{nm}
        ( c_{nm}^{\dag}  c_{nm} + d_{nm}^{\dag} d_{nm} + 1) +
        \frac{1}{2 \Im} v_0^{(1)2} +
        \frac{1}{M} v_{+1}^{(1)} v_{-1}^{(1)},
	\label{kk51}
\end{eqnarray}
where the sum over subindices $n$ include only finite frequency
modes.

The remaining terms in the hamiltonian are at most of ${\cal O}(f^{-1})$.
Therefore one might be tempted to use perturbation theory to evaluate
the corrections to the quadratic contributions. However, due to the
presence of the zero-modes such perturbative expansion would be plagued
with infrared divergences. As it will be shown in the next sections the
BRST quantization scheme provides a very convenient method to eliminate
these divergences in a consistent way.

\section{BRST quantization} \label{BRST}

The constraining operators $F_s\equiv v_s-V_s$ generate gauge
transformations, which are a manifestation of the fact that
transforming the fields and correspondingly moving the frame of
reference must  result into two completely equivalent physical
descriptions.

For each collective degree of freedom we introduce a Lagrange
multiplier $\Omega_s$ and two ghost operators $\eta_s, {\bar \eta}_s$,
together with the corresponding conjugate operators $B_s,\;\pi_s\;{\bar
\pi}_s$, which satisfy the non-vanishing commutation and
anti-commutation relations
\begin{equation}
{[} \Omega_s, B_{-t} ] = i \delta_{st} \; ; \;\;\;\;\;
	\{ \eta_s, \pi_{-t} \} =
	\{ \bar{\eta}_s, \bar{\pi}_{-t} \} = \delta_{st}.
\end{equation}

The quantal constraints
\begin{equation}
F_s|{\rm phys}\rangle = B_s|{\rm phys}\rangle = 0, \label{k52}
\end{equation}
on physical states are replaced by the requirement that physical states
should be annihilated by the BRST charge ${\cal Q}$ \cite{HT92}
\begin{equation}
{\cal Q} \equiv B_s \bar{\pi}_{-s} - F_s \eta_{-s} +
	\frac {i}{2} C_{st}^u \eta_{-s} \eta_{-t} \pi_u,
	\label{k53}
\end{equation}
where  $[v_s,v_t]=i C_{st}^u v_u$ and $[V_s,V_t]=-i C_{st}^u V_u$. The
non-vanishing terms of $C_{st}^u = -C_{ts}^u$ are
\begin{equation}
C_{0'+1}^{+1} = -C_{0' -1}^{-1} = C_{0 -1}^{-1} =
	-C_{0 +1}^{+1} = \frac{1}{2}. \label{ces}
\end{equation}

As a consequence of the fact that the charge ${\cal Q}$ is a nilpotent
and hermitian operator, we may add to the hamiltonian any term of the
form $\{\rho,{\cal Q}\}$, without altering the overlaps of the original
hamiltonian within the subspace annihilated by ${\cal Q}$.  We choose
\cite{BK90}
\begin{equation}
\rho= \Omega_{s'} \pi_{-s'} +
	\omega^2_{s'} \left( G_{s'}-\frac{1}{2\Im_{s'}}B_{s'} \right)
	\bar{\eta}_{-s'},
	\label{k54}
\end{equation}
which yields
\begin{eqnarray}
\{\rho,{\cal Q}\} &=& - \Omega_{s'} F_{-s'} +
	\omega_{s'}^2
	\left( G_{s'} B_{-s'} - \frac{1}{2\Im_{s'}} B_{s'} B_{-s'} \right) +
	i \pi_{s'}{\bar \pi}_{-s'} \nonumber \\
	&& +
	\omega_{s'}^2 \eta_{-t} \bar{\eta}_{-s'} [G_{s'},v_t] +
	i C_{s't}^u \Omega_{-s'} \eta_{-t} \pi_u.
	\label{k55}
\end{eqnarray}
The (spurious) frequencies $\omega_{s'}$ are arbitrary and should
disappear from any physical result. Since the collective variables are
to be considered as genuine variables of the problem, a tradeoff should
take place and some original degrees of freedom must become spurious.
At the quadratic level this is accomplished by including the zero modes
of the hamiltonian (\ref{kk51}) in the spurious sector, together with
the quadratic terms of $\{\rho,{\cal Q}\}$. Moreover, at this level
there is a separation of variables corresponding to each of the three
spurious sectors
\begin{eqnarray}
H^{(2)}_{\rm sp} &=& \frac{1}{2 \Im_{s'}} v^{(1)}_{s'} v^{(1)}_{-s'} -
	\Omega_{s'} v_{-s'}^{(1)} +
	\omega_{s'}^2 G_{s'} B_{-s'} -
	\frac{\omega_{s'}^2}{2\Im_{s'}} B_{s'} B_{-s'} +
	i \pi_{s'} \bar{\pi}_{-s'} +
	i \omega_{s'}^2 \eta_{s'} \bar{\eta}_{-s'} \nonumber \\
	&=&
	\omega_{s'}
	\left( a^{\dag}_{s'1} a_{s'1} - a^{\dag}_{s'0} a_{s'0} +
	\bar{a}_{s'} a_{s'} + \bar{b}_{s'} b_{s'}
	\right).
	\label{k56}
\end{eqnarray}
The normal boson and ghost modes may be obtained by applying the
transformations
\begin{eqnarray}
a_{s'1} &=& \sqrt{\frac{1}{2 \Im_{s'} \omega_{s'}}} v_{s'}^{(1)} -
	\sqrt{\frac{\Im_{s'}}{2 \omega_{s'}}} \Omega_{s'} -
	i \sqrt{\frac{\Im_{s'} \omega_{s'}}{2}} G_{s'}, \nonumber \\
a_{s'0} &=& \sqrt{\frac{\Im_{s'}}{2 \omega_{s'}}} \Omega_{s'} -
	i \sqrt{\frac{\omega_{s'}}{2 \Im_{s'}}} B_{s'} +
	i \sqrt{\frac{\Im_{s'} \omega_{s'}}{2}} G_{s'}, \nonumber \\
a_{s'} &=& -i \bar{b}_{-s'}^{\dag} =
	\sqrt{\frac{1}{2 \omega_{s'}}} \bar{\pi}_{s'} -
	i \sqrt{\frac{\omega_{s'}}{2}} \eta_{s'}, \nonumber \\
b_{s'} &=& i \bar{a}_{-s'}^{\dag} =
	\sqrt{\frac{1}{2 \omega_{s'}}} \pi_{s'} +
	i \sqrt{\frac{\omega_{s'}}{2}} \bar{\eta}_{s'}.
	\label{diag}
\end{eqnarray}

Note the manifest supersymmetry of the spurious sector and the
relations\footnote{The minus sign in the second commutation relation is
a consequence of the fact that we use the existing freedom to describe
the spurious sector by demanding that the vacuum state of the
$\Omega_s$-oscillators should be annihilated by the operator
$-\frac{\partial}{\partial \Omega_s}+\Omega_s$ \cite{BK90}.}
\begin{equation}
{[a_{s'1},a_{t'1}^{\dag}]}=-{[a_{s'0},a_{t'0}^{\dag}]}=
	\{a_{s'},{\bar a}_{t'}\} =
	\{b_{s'},{\bar b}_{t'}\} =\delta_{s't'}.
	\label{k58}
\end{equation}

The vacuum state of the spurious sector satisfies
\begin{equation}
a_{s'1}|\rangle=a_{s'0}|\rangle=a_{s'}|\rangle=b_{s'}|\rangle=0,
\end{equation}
it is also annihilated (to leading order) by the BRST charge, and it is
the only normalizable state of the spurious sector satisfying this
condition. It represents the unperturbed spurious sector for any
physical state.

We express the fields and their conjugate momenta in terms of the
corresponding eigenmodes and of the creation (annihilation) operators
for the finite frequency modes $c^{\dag}_{nm},d^{\dag}_{nm}$
$(c_{nm},d_{nm})$ and for the spurious modes
$a^{\dag}_{s'1},a^{\dag}_{s'0}$ $(a_{s'1},a_{s'0})$. In the Schroedinger
representation,
\begin{eqnarray}
\hat{\theta} &=& -\frac{1}{2} \sqrt{\frac{1}{2 \omega_0}}
	\Psi_0^{\rm ZM}(\vec{r})
	\left( a_{0,1} - a_{0,1}^{\dag} \right) -
	\frac{1}{2} \sqrt{\frac{1}{\omega_1}} \Psi_1^{\rm ZM}(\vec{r})
	\left( a_{1,1} - a_{-1,1}^{\dag} \right) \nonumber \\
        & & +\frac{1}{2} \sum_{n,m}
        \varepsilon_{nm}^{-1/2}
	\Psi_{nm}(\vec r)
	\left( c_{nm} + d_{nm}^{\dag} \right), \nonumber \\
\hat{\pi} &=& - i \sqrt{\frac{\omega_0}{2}}
	\left[ \Psi_0^{\rm ZM}(\vec{r}) \right]^*
	\left( a_{0,1} + a_{0,0} +
	a_{0,1}^{\dag} + a_{0,0}^{\dag} \right)
	\nonumber \\
	& & -
	i \sqrt{\frac{\omega_1}{2}}
	\left[ \Psi_1^{\rm ZM}(\vec{r}) \right]^*
	\left( a_{-1,1} + a_{-1,0} +
	a_{1,1}^{\dag} + a_{1,0}^{\dag} \right)
	\nonumber \\
	& & +
	i \sum_{n,m}
        \varepsilon_{nm}^{1/2}
	\Psi_{nm}^*(\vec r)
        \left( c_{nm}^{\dag} - d_{nm} \right),
	\label{modosn}
\end{eqnarray}
and the corresponding expressions for $\hat{\bar \theta}$ and
$\hat{\bar{\pi}}$. The normalized eigenmodes are defined
\begin{equation}
\Psi_{nm}(\vec r) = \sqrt{\frac{1}{2\pi}} R_{nm}(r) e^{i m\varphi},
\label{eigenmodes}
\end{equation}
for the finite modes, and
\begin{eqnarray}
\Psi^{\rm ZM}_0(\vec r) &=&
	\frac{2 f r  e^{i \varphi}}{\sqrt{\Im} (1+r^2)}, \nonumber \\
\Psi^{\rm ZM}_1(\vec r) &=&
	 - \frac{2 f}{\sqrt{M} (1+r^2)},
\label{zeromodes}
\end{eqnarray}
for the zero modes.

The residual BRST terms are
\begin{equation}
H_{\rm BRST}^{({\rm res})} = H^{({\rm res})} +
	\{\rho,{\cal Q}\}^{(3)},
\end{equation}
where
\begin{equation}
\{\rho,{\cal Q}\}^{(3)} = \Omega_{s'} V_{-s'} -
	\Omega_{s'} v^{(2)}_{-s'} +
	\omega_{s'}^{2} \eta_{-t} \bar{\eta}_{-s'} [G_{s'}, v_t^{(2)}] +
	i C_{s't}^u \Omega_{-s'} \eta_{-t} \pi_u, \label{CLIN}
\end{equation}
and we have used the fact that $v_{s'}$ is at most cuadratic in the
fluctuations.  Thus $\{\rho,{\cal Q}\}^{(3)}$ is of ${\cal O}(f^{-1})$.
As already mentioned the terms in $H^{({\rm res})}$ are also of ${\cal
O}(f^{-1})$ and higher.  In Eq.\ (\ref{CLIN}) there appear the ghost
conjugate operators $\eta_{0'},\pi_{0'}$ for which there is no
corresponding term in the quadratic hamiltonian (\ref{k56}). The
Hilbert space is thus divided into two degenerate subspaces by this
ghost degree of freedom. However, there are no IR-divergences, since
the ghost excitations always appear pairwise and the ghost accompanying
the $\eta_{0'},\pi_{0'}$ operators has a frequency $\omega_{s'}$.
Therefore, the full residual hamiltonian may be treated in perturbation
theory.  However arbitrary, the spurious frequencies $\omega_{s'}$ are
finite and therefore the problem of infrared divergences has
disappeared.  Counterterms should still be introduced in order to solve
the difficulties at the ultraviolet.

It should be noticed that, to ${\cal O}(f^{-1})$, the residual BRST
hamiltonian has only off-diagonal terms (cf.\ Eq.\ (\ref{h3})).
Therefore, the lowest order corrections to diagonal quantities (like
i.e., the mass of a soliton-meson state) turn out to be of ${\cal O}
(f^{-2})$.

The operator $L_{0'}$ defined as
\begin{eqnarray}
L_{0'} &=&
	-\{ {\cal Q}, \pi_{0'} + i C_{0's}^t \Omega_{-s} \bar{\eta}_t \}
	\nonumber \\
	&=& v_{0'} - V_{0'} - i C_{0' s}^{t} \Omega_{-s} B_t -
	C_{0' s}^t \eta_{-s} \pi_t -
	C_{0' s}^t \bar{\pi}_{-s} \bar{\eta}_t,
\end{eqnarray}
is the extension of $F_{0'}$ to include the transformation of the
Lagrange multipliers and ghosts. It commutes with $H_{\rm BRST}$ and
is diagonalized by the transformation (\ref{diag}),
\begin{equation}
L_{0'} = \Lambda - \frac{1}{2} (T_3 + J),
\end{equation}
where
\begin{eqnarray}
\Lambda &=& \sum_{nm} \frac{1}{2} (m-1)
	(c_{nm}^{\dag} c_{nm} - d_{nm}^{\dag} d_{nm} )
	\nonumber \\
	& & +
	\sum_{s'} C_{0' s'}^{s'}
	(a_{s'0}^{\dag} a_{s'0} - a_{s'1}^{\dag} a_{s'1} +
	\bar{a}_{s'} a_{s'} - \bar{b}_{s'} b_{s'} ). \label{lambda}
\end{eqnarray}

Let us consider as a basis of the Hilbert space the product form $|{\rm
intr} \rangle \otimes |{\rm coll} \rangle$. The intrinsic subspace is
characterized by the occupation numbers $n_{nm}^c,n_{nm}^d$
(=0,1,2,\ldots) of the real phonons and by the occupation numbers
$n_{s'0},n_{s'1}$ (=0,1,2,\ldots) and $n_{s'a},n_{s'b}$ (=0,1) of the
spurious phonons and ghosts. It carries the quantum number $\Lambda$.

A complete set of states for the collective sector depends on the angle
$\alpha$ corresponding to internal rotation, the angle $\Phi$
determining the orientation of the spatial frame, and the coordinates
$Z,\bar{Z}$ associated with the translational motion in the plane. It
is convenient to use the complete set given by
\begin{equation}
\vert  T_3 , J , P_L , \bar{P}_L \rangle =
	e^{i T_3 \alpha} e^{i J \Phi}
	e^{i (P_L Z + \bar{P}_L \bar{Z})}.
\end{equation}

The operator $L_{0'}$ annihilates physical states since it is a
``null'' operator \cite{BK90}. Therefore, for such states $T_3 = 2
\Lambda - J$ and the collective subspace of interest is of the form (up
to a trivial phase)
\begin{equation}
\vert  \Lambda ; J , P_L , \bar{P}_L \rangle =
	e^{i J (\Phi - \alpha)}
	e^{i (P_L Z + \bar{P}_L \bar{Z})}, \label{cost}
\end{equation}
where $\Lambda$ is determined from the intrinsic structure.

The collective-intrinsic coupling (${\cal O}(f^{-1})$) is given by the
first term in Eq.\ (\ref{CLIN}),
\begin{equation}
\Omega_{s'} V_{-s'} = \Omega_0 (\Lambda - J) +
	i \sqrt{2} \left(
	e^{-i \Phi} \Omega_1 \bar{P}_L -
	e^{i \Phi} \Omega_{-1} P_L \right).
\end{equation}
The operator $\Omega_0$ does not change the value of $\Lambda$, while
$\Omega_1$ ($\Omega_{-1}$) decreases (increases) the value of $\Lambda$
by one unit (cf. Eqs.\ (\ref{ces}), (\ref{diag}) and (\ref{lambda})).
The conservation of $L_{0'}$ is insured by the terms $e^{\mp i \Phi}$
changing the value of $J$ by the necessary amount.

The collective parameters may be calculated in perturbation theory
using the fact that the intrinsic energies are of ${\cal O}(1)$ and
thus much larger than the collective energies (${\cal O}(f^{-2})$,
cf.\ Eq.\ (\ref{k60})). In second order of perturbation theory such
term yields the well-known collective energies
\begin{equation}
H_{coll} = \frac{1}{2 \Im_{s'}} V_{s'} V_{-s'} =
	\frac{1}{2 \Im} (J - \Lambda)^2 +
	\frac{2}{M} P_L \bar{P}_L,
	\label{k60}
\end{equation}
where Eqs.\ (\ref{pl}) and (\ref{kk51}) have been used. The collective
energies are given by the expectation values of (\ref{k60}) within the
subset of states (\ref{cost}) associated with given values of
$\Lambda$, $J$,  $P_L$ and $\bar{P}_L$.

Higher orders of perturbation theory yield corrections to the
parameters $\Im$ and $M$, as well as higher than quadratic terms in the
collective operators.

\section{Correction to the soliton mass} \label{DIAGS}

As an application of the formalism developed above in this section we
evaluate explicitly the correction of ${\cal O}(f^{-2})$ to the soliton
mass.  For convenience we work with the real fields $\theta_a$
($a=1,2$) which are related to the fields $\theta$ and $\bar{\theta}$
of the previous sections by
\begin{equation}
\theta = \theta_1 + i \theta_2 \; ; \;\;\;\;\;
	\bar{\theta} = \theta_1 - i \theta_2
\end{equation}

\begin{figure}
\begin{picture}(36000,12000)
\drawline\fermion[\N \REG](12000,4500)[3000]
\global\advance \pfronty by -1500
\global\advance \pbacky by 1500
\put(\pfrontx,\pfronty){\circle{3000}}
\put(\pbackx,\pbacky){\circle{3000}}
\global\advance \pfronty by 1400
\global\advance \pbacky by -1400
\put(\pfrontx,\pfronty){\circle*{300}}
\put(\pbackx,\pbacky){\circle*{300}}
\global\advance \pbackx by -750
\put(\pbackx,0){(a)}
\bezier{200}(20000,2000)(16000,6000)(20000,10000)
\bezier{200}(20000,2000)(24000,6000)(20000,10000)
\drawline\fermion[\N \REG](20000,2000)[8000]
\put(\pfrontx,\pfronty){\circle*{300}}
\put(\pbackx,\pbacky){\circle*{300}}
\global\advance \pbackx by -750
\put(\pbackx,0){(b)}
\put(28000,8000){\circle{4000}}
\put(28000,4000){\circle{4000}}
\put(28000,6000){\circle*{300}}
\put(27250,0){(c)}
\end{picture}
\caption{Diagrammatic corrections of ${\cal O}(f^{-2})$ to the soliton
mass.} \label{diags}

\end{figure}
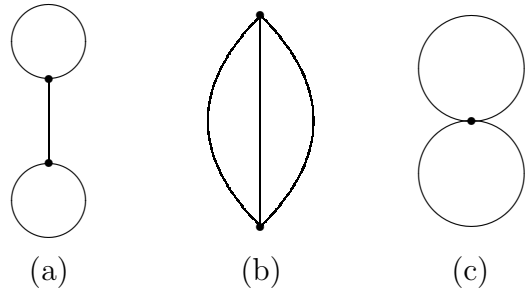

The correction of ${\cal O}(f^{-2})$ to the soliton mass is given by
two-loops vacuum diagrams (Fig.\ \ref{diags}) obtained from $H_{\rm
BRST}^{(3)}$ and $H_{\rm BRST}^{(4)}$.
\begin{eqnarray}
H_{\rm BRST}^{(3)} &=&
	\frac{1}{6 f} \int \left\{
	3 x_a \hat{\theta}_a \hat{\pi}_b \hat{\pi}_b +
	4 (1+r^2)^3 G_{abc} \hat{\theta}_a \hat{\theta}_b \hat{\theta}_c +
	12 (1+r^2)^2 G_{bc}
	\partial_a \left[
	(1+r^2) \hat{\theta}_a \right] \hat{\theta}_b \hat{\theta}_c
	\right.
	\nonumber \\ & & + \left.
	12 G_b
	\nabla \left[ (1+r^2) \hat{\theta}_a \right] \cdot
	\nabla \left[
	(1+r^2) \hat{\theta}_a \right] \hat{\theta}_b \right\} +
	\{\rho,{\cal Q}\}^{(3)}, \label{h3} \\
H_{\rm BRST}^{(4)} &=&
	\frac{1}{24 f^2} \int \left\{
	12 x_a x_b \hat{\theta}_a \hat{\theta}_b \hat{\pi}_c \hat{\pi}_c +
	6 (1+r^2) \hat{\theta}_a \hat{\theta}_a \hat{\pi}_b \hat{\pi}_b +
	4 (1+r^2)^4 G_{abcd} \hat{\theta}_a \hat{\theta}_b
	\hat{\theta}_c \hat{\theta}_d
	\right. \nonumber \\ & & +
	16 (1+r^2)^3 G_{bcd}
	\partial_a \left[ (1+r^2) \hat{\theta}_a \right]
	\hat{\theta}_b \hat{\theta}_c \hat{\theta}_d
	\nonumber \\ & & + \left.
	24 (1+r^2)^2 G_{bc}
	\nabla \left[ (1+r^2) \hat{\theta}_a \right] \cdot
	\nabla \left[ (1+r^2) \hat{\theta}_a \right]
	\hat{\theta}_b \hat{\theta}_c \right\},
\end{eqnarray}
where
\begin{equation}
G_{a_1...a_k} = \left. \frac{\partial^k}{\partial_{a_1}...\partial_{a_k}}
	(1 + \theta^2)^{-2} \right|_{\theta_a=x_a}
\label{Gdef}
\end{equation}
and $\{\rho,{\cal Q}\}^{(3)}$ is given in (\ref{CLIN}).

Each diagram of Fig.\ \ref{diags} has terms which depend on the
spurious frequencies. They are
\begin{eqnarray}
\mbox{(a)} &=& - \frac{1}{64} \sum_{n_1,n_2,s'}
	\frac{1}{\omega_{n_1} \omega_{n_2}^2 \omega_{s'}}
	A_{s' s' n_2} \left(A_{n_1 n_1 n_2} -
	2 \omega_{n_1}^2 B_{n_1 n_1;n_2}\right)
	\nonumber \\ & & -
	\frac{1}{128} \sum_{n,s',t'}
	\frac{1}{\omega_{n}^2 \omega_{s'} \omega_{t'}}
	A_{s' s' n} A_{t' t' n}, \\
\mbox{(b)} &=& \frac{1}{8} \sum_{n_1,n_2,s'}
	\frac{\omega_{n_1} - \omega_{n_2}}
	{\Im_{s'} \omega_{n_2} \omega_{s'}}
	\left[
	2 \sqrt{\Im_{s'}} \omega_{s'} E_{s';n_1;n_2} B_{n_1 s';n_2} -
	(\omega_{n_1} + \omega_{n_2} - \omega_{s'})
	E_{s';n_1;n_2}^2 \right]
	\nonumber \\ & & +
	\frac{1}{16} \sum_{n,s',t'} \frac{1}{\Im_{s'}} \left[
	E_{s';t';n}
	\left( \frac{\omega_n^2}{\omega_{s'} \omega_{t'}} -
	2 \frac{\omega_n}{\omega_{s'}} + 2 \right) -
	2 \sqrt{\Im_{s'}} E_{s';t';n} B_{s' t';n} \right]
	\nonumber \\ & & -
	\frac{1}{288} \sum_{s',t',u'} \left(
	\frac{\Im_{u'}}{\Im_{s'} \Im_{t'}}
	C_{s' t'}^{u'} C_{s' t'}^{u'} +
	\frac{2}{\Im_{s'}} C_{s' t'}^{u'} C_{s' u'}^{t'}
	\right), \\
\mbox{(c)} &=& \frac{1}{32} \sum_{n,s'}
	\frac{1}{\omega_{n} \omega_{s'}} \left(
	F_{n n s' s'} - 2 \omega_{n}^2 D_{n n;s' s'} -
	8 D_{n s';n s'} \right)
	\nonumber \\ & & +
	\frac{1}{64}
	\sum_{s',t'} \frac{1}{\omega_{s'} \omega_{t'}} \left(
	F_{s' s' t' t'} - 8 D_{s' t';s' t'} \right).
\end{eqnarray}
The integrals $A$, $B$, $D$, $E$ and $F$ are defined in the Appendix.
The indices $n_k$ run only over finite frequency modes. Indices
$s'$,$t'$,$u'$ label the zero-modes.

Summing up the expressions $(a)$, $(b)$ and $(c)$ and
including the real sector contributions we obtain the
total correction of order ${\cal O}(f^{-2})$ to the soliton mass.
It reads:

\begin{eqnarray}
\Delta M^{(2)} &=& - \frac{1}{192} \sum_{n_1,n_2,n_3}
	\frac{
	\left(A_{n_1 n_2 n_3} +
	2 \omega_{n_1} \omega_{n_2} B_{n_1 n_2;n_3} +
	2 \omega_{n_3} \omega_{n_1} B_{n_3 n_1;n_2} +
	2 \omega_{n_2} \omega_{n_3} B_{n_2 n_3;n_1}
	\right)^2}
	{\omega_{n_1} \omega_{n_2} \omega_{n_3}
	(\omega_{n_1} + \omega_{n_2} + \omega_{n_3})}
	\nonumber \\ & &  - \frac{1}{128} \sum_{n_1,n_2,n_3}
	\frac{1}{\omega_{n_1} \omega_{n_2} \omega_{n_3}^2}
	\left(A_{n_1 n_1 n_3} - 2 \omega_{n_1}^2 B_{n_1 n_1;n_3}\right)
	\left(A_{n_2 n_2 n_3} - 2 \omega_{n_2}^2 B_{n_1 n_1;n_3}\right)
	\nonumber \\ & & +
	\frac{1}{8} \sum_{n_1,n_2,n_3}
	\frac{\omega_{n_1}}{\omega_{n_2}}
	B_{n_1 n_3;n_2}^2 +
 	\frac{1}{64} \sum_{n_1,n_2} \left(
	\frac{1}{\omega_{n_1} \omega_{n_2}}
	F_{n_1 n_1 n_2 n_2} -
	\frac{4 \omega_{n_1}}{\omega_{n_2}} D_{n_1 n_1;n_2 n_2}
	\right)
	\nonumber \\ & & +
	\frac{1}{8} \sum_{n_1,n_2,s'}
	\frac{\omega_{n_1} - \omega_{n_2}}{\Im_{s'} \omega_{n_2}}
	E_{s';n_1;n_2} \left( E_{s';n_1;n_2} -
	2 \sqrt{\Im_{s'}} B_{s' n_1;n_2} \right)
	\nonumber \\ & & +
	\frac{1}{8} \sum_{n,s',t'}
	\frac{1}{\Im_{s'}} E_{s';t';n}
	\left( E_{s';t';n} - \sqrt{\Im_{s'}} B_{s' t';n} \right)
	\nonumber \\ & & -
	\frac{1}{288} \sum_{s',t',u'} \left(
	\frac{\Im_{u'}}{\Im_{s'} \Im_{t'}}
	C_{s' t'}^{u'} C_{s' t'}^{u'} +
	\frac{2}{\Im_{s'}} C_{s' t'}^{u'} C_{s' u'}^{t'}
	\right).
\end{eqnarray}

Using the identities (\ref{id}) listed in the Appendix
it can be explicitly checked that $\Delta M^{(2)}$  is
independent of the arbitrary parameters $\omega_{s'}$, as it should.

\section{Conclusions} \label{CON}

We have carried out a quantum mechanical treatment of the $O(3)$ model
in (2+1) dimensions, both for the collective and the intrinsic
excitations. We have derived the explicit expression of the BRST
hamiltonian and shown that the corrections to the quadratic expressions
may be perturbatively performed in terms of an expansion in the inverse
of the large parameter of the model $f$. Such expansion is free of
IR-divergences related to translations and rotations of the classical
solution. There still remains the IR-divergences due to massless
mesons which are already present in the vacuum sector and can be regulated
 including a mass term.

The lagrangian we have studied is not UV-renormalizable. It must be
considered as the low energy limit of a lagrangian with an infinite
number of higher order derivative terms which can absorb the
divergences \cite{We79}.

Higher order terms must also be taken into account to avoid the
dilatational instability. However, in the present work we have
preferred not to include them and ignore the dilatation problem
altogether. In this way we have the simplicity associated with the
existence of analytical solutions and keep most of the features
inherent to soliton models. Namely, there is a set of several
zero-modes associated with broken symmetries, a non-commutative algebra
associated such symmetries, the interplay between external and internal
transformations and the survival of some unbroken symmetries.  We have
shown that such problems may be successfully treated by the application
of methods based on the BRST invariance. Moreover, such procedure may
be carried out to other more realistic models in a fairly
straightforward way, but for the fact that the equations should be
numerically solved.

In our formalism we have considered that the collective velocities are
small, i.e., $|V_{s'}|/\Im_{s'} \ll 1$.  This allows to treat
perturbatively the intrinsic-collective coupling $\Omega_{s'}V_{s'}$.
In principle, the BRST procedure allows also to treat cases in which
some expectation values of the Lagrange multipliers $\langle
\Omega_{s'} \rangle = \langle V_{s'} \rangle / \Im_{s'}$ are of ${\cal
O}(1)$. Such treatment has been applied, for instance, to the
translational motion in \cite{ABS93} and corresponds to the classical
solution used in \cite{DHM94} for the internal motion (in both cases
(1+1) dimensions is assumed). An obvious and interesting extension of
the present work would be the treatment of models with non-commutative
zero-modes in the large collective momentum limit. This would allow,
among other things, to study the stability of the fast rotating
solitons. Another point that could be studied is the appearance of the
long-time-seek Yukawa coupling in such a limit as it has been recently
suggested in Ref.\ \cite{DHM94}. We hope to be able to report on these
topics in the near future.

\section*{Acknowledgments}

This work was performed with the support of Fundaci\'on Antorchas.  D.\
R.\ B.\ acknowledges the hospitality of the Theoretical Physics Institute
at the University of Minnesota, where this work was partially finished
under DOE Nuclear Contract No.\ DE-FG0287ER-40328.

\appendix

\section*{}

In Sec.\ \ref{DIAGS} we made use of the following integrals which
appear in the vertices of $H_{\rm BRST}^{(3)}$ and $H_{\rm
BRST}^{(4)}$
\begin{eqnarray}
A_{\bar{n}_1 \bar{n}_2 \bar{n}_3} &=& \frac{1}{6 f} \int \left\{
	(1+r^2)^3 G_{abc} \psi_{a,\bar{n}_1} \psi_{b,\bar{n}_2}
	\psi_{c,\bar{n}_3} \right.
	\nonumber \\ & & +
	3 (1+r^2)^2 G_{bc}
	\partial_a \left[ (1+r^2) \psi_{a,\bar{n}_1} \right]
	\psi_{b,\bar{n}_2} \psi_{c,\bar{n}_3}
	\nonumber \\ & & + \left.
	3 (1+r^2) G_{b}
	\nabla \left[ (1+r^2) \psi_{a,\bar{n}_1} \right] \cdot
	\nabla \left[ (1+r^2) \psi_{a,\bar{n}_2} \right]
	\psi_{b,\bar{n}_3} \right\}
	\nonumber \\ & & +
	\mbox{permutations of\ }(\bar{n}_1,\bar{n}_2,\bar{n}_3), \\
B_{\bar{n}_1 \bar{n}_2;\bar{n}_3} &=& \frac{1}{2f} \int
	(1+r^2)^3 G_b \psi_{a,\bar{n}_1}
	\psi_{a,\bar{n}_2} \psi_{b,\bar{n}_3}, \\
D_{\bar{n}_1 \bar{n}_2;\bar{n}_3 \bar{n}_4} &=& \frac{1}{4f^2} \int
	(1+r^2)^4 G_{bc} \psi_{a,\bar{n}_1}
	\psi_{a,\bar{n}_2} \psi_{b,\bar{n}_3} \psi_{c,\bar{n}_4}, \\
E_{0;\bar{n}_1;\bar{n}_2} &=& 0, \\
E_{\pm 1;\bar{n}_1;\bar{n}_2} &=& \pm \frac{i}{4 \sqrt{2} f^2}
	\int (1+r^2) \psi_{a,\bar{n}_1}
	(\partial_1 \mp i \partial_2)
	\left[ (1+r^2) \psi_{a,\bar{n}_2} \right], \\
F_{\bar{n}_1 \bar{n}_2 \bar{n}_3 \bar{n}_4} &=&
	\frac{1}{48 f^2} \int \left\{
	(1+r^2)^4 G_{abcd} \psi_{a,\bar{n}_1} \psi_{b,\bar{n}_2}
	\psi_{c,\bar{n}_3} \psi_{d,\bar{n}_4} \right.
	\nonumber \\ & & +
	4 (1+r^2)^3 G_{bcd}
	\partial_a \left[ (1+r^2) \psi_{a,\bar{n}_1} \right]
	\psi_{b,\bar{n}_2} \psi_{c,\bar{n}_3} \psi_{d,\bar{n}_4}
	\nonumber \\ & & + \left.
	6 (1+r^2)^2 G_{bc}
	\nabla \left[ (1+r^2) \psi_{a,\bar{n}_1} \right] \cdot
	\nabla \left[ (1+r^2) \psi_{a,\bar{n}_2} \right]
	\psi_{b,\bar{n}_3} \psi_{c,\bar{n}_4} \right\}
	\nonumber \\ & & +
	\mbox{permutations of\ }(\bar{n}_1,\bar{n}_2,\bar{n}_3,\bar{n}_4),
\end{eqnarray}
where $G_{a_1...a_k}$ is defined in Eq.\ (\ref{Gdef}), and
$\psi_{a,\bar{n}}$ are the eigenmodes of the fluctuation (see
Eqs.\ (\ref{eigenmodes},\ref{zeromodes})), which read
\begin{equation}
\psi_{1,n m}(\vec{r}) =
	\frac{1}{2 \sqrt{\pi}} R_{n m}(r) \cos{m \varphi}
	\; ; \;\;\;\;\;
	\psi_{2,n m}(\vec{r}) =
	\frac{1}{2 \sqrt{\pi}} R_{n m}(r) \sin{m \varphi},
\end{equation}
for the finite frequency modes, and
\begin{eqnarray}
\psi_{1,0}^{\rm ZM}(\vec{r}) &=&
	\sqrt{\frac{2}{\Im}} \frac{f x_1}{(1+r^2)}
	\; ; \;\;\;\;\;
	\psi_{2,0}^{\rm ZM}(\vec{r}) =
	\sqrt{\frac{2}{\Im}} \frac{f x_2}{(1+r^2)}, \\
\psi_{1,\pm 1}^{\rm ZM}(\vec{r}) &=&
	\mp \sqrt{\frac{2}{M}} \frac{f}{(1+r^2)}
	\; ; \;\;\;\;\;
	\psi_{2,\pm 1}^{\rm ZM}(\vec{r}) = 0,
\end{eqnarray}
for the zero-modes.

We list below the relations that the integrals $A$, $B$, $E$ and $F$
satisfy when some of the indices $\bar{n}$ correspond to zero-modes:
\begin{eqnarray}
A_{s' n_1 n_2} &=& - \frac{2}{\sqrt{\Im_{s'}}}
	\left( \omega_{n_1}^2 - \omega_{n_2}^2 \right)
	E_{s';n_1;n_2}, \nonumber \\
A_{s' t' n} &=& \frac{2}{\sqrt{\Im_{s'}}} \omega_{n}^2
	E_{s';t';n}, \nonumber \\
A_{s' t' u'} &=& 0, \nonumber \\
B_{n_1 n_2;s'} &=& 0 \label{id}\\
F_{s' s' n_1 n_1} &=& \frac{1}{\sqrt{\Im_{s'}}}
	\sum_{n_2} E_{s';s';n_2}
	\left( A_{n_1 n_1 n_2} -
	2 \omega_{n_1}^2 B_{n_1 n_1;n_2} \right)
	\nonumber \\ & &
	+ \frac{4}{\Im_{s'}} \sum_{n_2}
	\left( \omega_{n_1}^2 - \omega_{n_2}^2 \right)
	E_{s';n_1;n_2}^2 - \frac{4}{\Im_{s'}} \sum_{t'}
	\omega_{n_1}^2 E_{s';t';n_1}^2, \nonumber \\
F_{s' s' t' t'} &=& \frac{2}{\sqrt{\Im_{s'} \Im_{t'}}} \sum_{n_1}
	E_{s';s';n_1} E_{t';t';n_1} +
	\frac{4}{\Im_{s'}} \sum_{n_1} \omega_{n_1}^2
	E_{s';t';n_1}^2. \nonumber
\end{eqnarray}
$n_k$ denote finite frequency modes, and $s'$, $t'$ and $u'$
zero-modes. The above Eqs.\ relate the spurious vertices from $H^{(3)}$
and $H^{(4)}$ to the vertices from $\{\rho,{\cal Q}\}^{(i)}$, and are
needed to show that the total contribution from diagrams of ${\cal
O}(f^{-2})$ is independent of the spurious frequencies.

\end{document}